\documentclass[%
 reprint,
superscriptaddress,
 amsmath,amssymb,
 aps,
 longbibliography,
prb
]{revtex4-2}

\usepackage{graphicx}
\usepackage{dcolumn}
\usepackage{bm}
\usepackage{xcolor}
\usepackage{times}
\usepackage[normalem]{ulem}
\usepackage{units}
\usepackage[hypertexnames=false,linktocpage=true,colorlinks=true,linkcolor=blue,anchorcolor=blue,citecolor=blue,filecolor=blue,urlcolor=blue,bookmarksnumbered=true,pdfview=FitB,breaklinks=true]{hyperref}

\begin{document}
\title{Structural and magnetic properties of TbCuAs$_2$ studied by X-ray and neutron scattering}

\author{M.~G.~Kim}\email{mgkim@uwm.edu}
\affiliation{Department of Physics, University of Wisconsin-Milwaukee, Milwaukee, WI 53201, USA}

\author{T. Heitmann}
\affiliation{University of Missouri Research Reactor, University of Missouri, Columbia, MO 65211, USA}
\affiliation{Department of Physics and Astronomy, University of Missouri, Columbia, MO 65211, USA}
\affiliation{Materials Science and Engineering Institute, University of Missouri, Columbia, MO 65211, USA}

\author{S.~Boney}
\affiliation{Department of Physics, University of Wisconsin-Milwaukee, Milwaukee, WI 53201, USA}

\author{C.~Neupane}
\affiliation{Department of Physics, University of Wisconsin-Milwaukee, Milwaukee, WI 53201, USA}

\author{R. Acevedo-Esteves}
\affiliation{National Synchrotron Light Source II, Brookhaven National Laboratory, Upton, New York 11973, USA}

\author{A. Sapkota}
\affiliation{Ames National Laboratory, Iowa State University, Ames, Iowa 50011, USA}

\author{D. Evans}
\affiliation{Department of Physics, Simon Fraser University, Burnaby, British Columbia, Canada}

\author{P. C. Canfield}
\affiliation{Ames National Laboratory, Iowa State University, Ames, Iowa 50011, USA}
\affiliation{Department of Physics and Astronomy, Iowa State University, Ames, Iowa 50011, USA}

\author{C. Nelson}
\affiliation{National Synchrotron Light Source II, Brookhaven National Laboratory, Upton, New York 11973, USA}

\author{E.~D.~Mun}
\affiliation{Department of Physics, Simon Fraser University, Burnaby, British Columbia, Canada}

\author{J.-W. Kim}
\affiliation{Advanced Photon Source, Argonne National Laboratory, Argonne, Illinois 60439, USA}

\date{\today}


\begin{abstract}

We investigated antiferromagnetic metal TbCuAs$_2$, which exhibits an anomalous resistivity upturn at low temperatures. This resistivity upturn has been proposed to originate from the material's magnetic ground state; however, the precise nature of the antiferromagnetic ordering in this compound has not been previously established. Here, we present a combined single crystal neutron diffraction, X-ray resonant magnetic scattering, and single crystal X-ray diffraction study of TbCuAs$_2$. These measurements unambiguously determine the Tb magnetic moment direction, the antiferromagnetic arrangement along high symmetry directions, the interplay between structure and magnetism, and their temperature dependence. Our observations are compared with previous studies on sister compounds in the rare-earth copper arsenide family ($R$CuAs$_2$, $R$ = Sm and Gd), which also displays resistivity anomalies. We suggest a common underlying physics governing the structural and magentic properties of rare-earth copper arsenides that exhibit low-temperature resistivity anomalies.

\end{abstract}

\maketitle

\section{introduction}
In a conventional metal, the electrical resistivity decreases upon cooling as phonon scattering is progressively suppressed.\cite{Metals-1,metals-2}  However, some metallic systems exhibit the opposite behavior at low temperatures, developing a resistivity minimum followed by an anomalous resistivity upturn before entering an ordered ground state. This unusual transport response signals the presence of an additional low-energy electron-scattering mechanism that becomes increasingly important as thermal fluctuations are reduced. While such behavior is classically associated with the Kondo effect in dilute magnetic alloys,~\cite{kondo-1,kondo-3, MAPLE-1,kondo-2} similar low-temperature resistivity upturns have also been observed in systems where a conventional Kondo interpretation is unlikely,~\cite{EDELSTEIN1968614,concentrated-magnet-1,PhysRevB.29.1088,PhysRevB.43.6042,matallic-glass-1,metallic-glass-2,bulk-ceramic-1,epi-film-1,epi-film-2,films-1,films-2} indicating that alternative mechanisms must be considered.

The rare-earth copper arsenide family $R$CuAs$_2$ ($R$ = rare earth) provides an appealing platform for studying this problem.\cite{Sampathkumaran-2003,SENGUPTA2004465, Evans,Ashiwini,Kim-1} Several members of this family, including $R$ = Nd, Sm, Gd, Tb, and Dy, exhibit a pronounced low-temperature resistivity upturn prior to the onset of long-range antiferromagnetic order, whereas others such as PrCuAs$_2$ do not.\cite{Sampathkumaran-2003,SENGUPTA2004465, Evans} The selective appearance of this transport anomaly within the same structural family strongly suggests that it is tied to the details of the magnetic ground state rather than to a trivial chemical trend. In particular, the emergence of the resistivity upturn just above the ordering temperature points to enhanced electron scattering from short-range magnetic correlations or fluctuating spin textures that develop before static order is fully established.

A compelling possibility is that the anomalous transport in these compounds originates from a magnetically frustrated regime in which competing exchange interactions generate correlated but only partially ordered spin states. In such a scenario, conduction electrons can be strongly scattered by spatially modulated magnetic correlations, especially when the magnetic structure is incommensurate or when substantial spectral weight develops at finite wave vectors. Recent theoretical work has suggested that frustration-driven liquid-like spin correlations in local-moment metals can enhance electron backscattering and produce a low-temperature resistivity minimum even in the absence of Kondo physics.~\cite{Wang-2016} From this perspective, the resistivity upturn in selected $R$CuAs$_2$ compounds may reflect an electronically visible precursor state to antiferromagnetic order, governed by the wave-vector-dependent spin correlations of the rare-earth moments.

Understanding this behavior requires knowledge not only of the magnetic structure, but also of the crystal structure and their coupled evolution with temperature. In this family, subtle structural distortions may play an essential role by modifying exchange pathways, lifting magnetic degeneracies, and selecting particular ordering wave vectors. Earlier studies established that $R$CuAs$_2$ compounds crystallize in a tetragonal structure at room temperature\cite{Mozharivskyj-2000,Mozharivskyj-2002,JEMETIO200293,Sampathkumaran-2003,SENGUPTA2004465}, but more recent investigations have shown that symmetry lowering can occur in at least some members of the family at low temperatures.\cite{Ashiwini} Such structural changes are particularly important in systems with frustrated or nearly degenerate magnetic interactions, where even weak lattice distortions can reorganize the magnetic energy landscape and stabilize complex ordered states. The low-temperature electrical behavior may therefore be governed not simply by magnetism alone, but by the coupled evolution of lattice symmetry, magnetic frustration, and conduction-electron scattering.

Within this family, TbCuAs$_2$ is especially compelling because it combines a pronounced resistivity upturn with a reported incommensurate antiferromagnetic ground state. Earlier neutron powder diffraction studies identified multiple incommensurate propagation vectors in TbCuAs$_2$,\cite{Zhao-2017}  implying that its magnetic ordering is substantially more complex than the relatively simpler antiferromagnetic structures reported in some sister compounds. Such incommensurate order is a natural fingerprint of competing interactions and possible magnetic frustration, and it raises the possibility that the same underlying instability responsible for the complex magnetic state also governs the anomalous transport.\cite{Kim-1} However, the exact magnetic structure of TbCuAs$_2$, including the Tb moment direction and detailed spin arrangement, has not been established. Moreover, the possible role of low-temperature structural symmetry lowering and its connection to the incommensurate magnetic order remain unresolved.

In this work, we investigate TbCuAs$_2$ using a combination of single-crystal neutron diffraction, X-ray resonant magnetic scattering, and single-crystal X-ray diffraction. By resolving both the magnetic and crystallographic structures and tracking their temperature dependence, we aim to determine how magnetic frustration, structural symmetry, and electron scattering are intertwined in this material. Our results show that the low-temperature state of TbCuAs$_2$ cannot be understood as a simple antiferromagnet, but instead emerges from a coupled structural and magnetic instability. These findings provide a microscopic framework for understanding the strange metallic transport observed in TbCuAs$_2$ and offer broader insight into the origin of resistivity upturns in the $R$CuAs$_2$ family.

\section{experiments}
Single crystals of TbCuAs$_2$ were grown out of a melt with excess Cu and As.~\cite{Evans,Canfield-1992,Ashiwini} The constituent elements of high-purity Tb, Cu, and As, taken in the ratio Tb$_{0.04}$(Cu$_{0.5}$As$_{0.5}$)$_{0.96}$, were loaded in an alumina crucible and sealed in a silica ampoule under partial argon pressure. The ampoule was heated slowly to 1050$^\circ$C and cooled down to 800 $^\circ$C over 120 hours. After removing the excess liquid by centrifuging,\cite{Canfield_2020} shiny single crystals were obtained. The phase purity of crystals from each growth batch was verified using room-temperature X-ray powder diffraction on a Rigaku Miniflex diffractometer. The magnetic and electrical properties of the as-grown single crystals were characterized using a Quantum Design Magnetic Property Measurement System and a Physical Property Measurement System, ensuring consistency with the expected bulk behavior prior to detailed scattering studies.

Temperature-dependent X-ray resonant magnetic scattering (XRMS) measurements were carried out on a six-circle diffractometer at beamline 6-ID-B of the Advanced Photon Source at Argonne National Laboratory. All measurements were performed at the Tb $L_3$ absorption edge ($E = 7.517$ keV) to enhance sensitivity to the Tb $4f$ magnetic moments. Thus, the XRMS measurements selectively probe the magnetic moments of the Tb ions. A plate-like single crystal was mounted on a flat copper sample holder attached to the cold finger of a closed-cycle Joule–Thomson cryostat, providing a base temperature of approximately 2 K. The sample was initially aligned with the tetragonal [1, 1, 0] direction in the scattering plane, and the azimuthal angle was controlled using the six-circle geometry to access different crystallographic orientations, including configurations with the tetragonal [0, 1, 0] in the scattering plane. We use the tetragonal notation for Bragg reflections throughout the paper.

The incident X-ray beam was linearly polarized perpendicular to the vertical scattering plane ($\sigma$ polarization). Under these conditions, dipole resonant magnetic scattering rotates the scattered polarization into the scattering plane ($\pi$ polarization), allowing for a direct determination of the magnetic moment orientation. Only the components of the magnetic moment projected onto the scattering plane contribute to the scattered intensity, while components perpendicular to the plane do not contribute. Scattered intensities were recorded using a two-dimensional detector, enabling efficient mapping of reciprocal space and magnetic reflections.

High-resolution, temperature-dependent single-crystal X-ray diffraction measurements were conducted on a six-circle diffractometer at 6-ID-B using a liquid-helium-flow closed-cycle refrigerator with a base temperature of approximately 6 K. The plate-like single crystal was mounted on a copper sample holder with the diffractometer $\Phi$-axis oriented approximately along the surface normal, coprresponding to the crystallographic \textbf{\textit{c}} axis. The crystal quality was confirmed by a mosaicity of less than $0.02^\circ$ full width at half maximum, determined from the rocking curve of the (0,~0,~4) reflection at room temperature. Diffraction data were collected as a function of temperature from room temperature down to the base temperature of the cryostat.

Single-crystal neutron diffraction measurements were performed using the TRIAX triple-axis spectrometer at the University of Missouri Research Reactor. The instrument was configured with beam collimations of $60'-40'-~$sample$~-40'-80'$, and measurements were carried out with fixed incident and final energies of $E_\mathrm{i} = E_\mathrm{f} = 14.7$ meV. Two pyrolytic graphite filters, placed before the monochromator and analyzer, were used to suppress higher-order neutron contamination. The single crystal was aligned in the $(H, 0, L)$ scattering plane and mounted in a closed-cycle refrigerator. Diffraction measurements were performed over a temperature range extending from room temperature down to approximately 5 K. Through this report, error bars represent one standard deviation.

\begin{figure}[!t]
    \centering
    \includegraphics[width=1\linewidth]{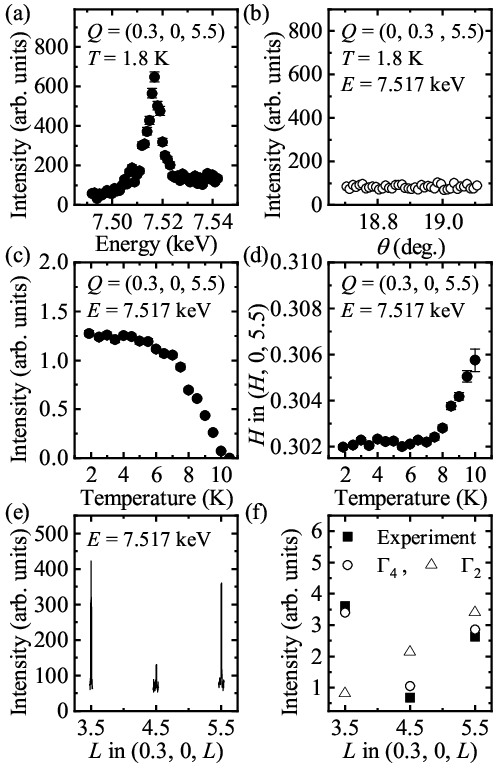}\\
    \caption{Resonant magnetic scattering from the TbCuAs$_2$ single crystal. Energy scan through (a) \textbf{\textit{Q}} $=$ (0.3,~0,~5.5) and (b) (0,~0.3,~5.5) magnetic peak positions at $T =$~1.8~K. (c) AFM order parameter measured at \textbf{\textit{Q}} $=$ (0.3,~0,~5.5) at the Tb $L_3$ edge ($E =$ 7.517 keV). (d) Change in the magnetic peak position measured at \textbf{\textit{Q}} $=$ (0.3,~0,~5.5). (e) Peak intensity at different $L$ in (0.3,~0, $L$). (f) Comparison between the measured intensity and the calculated intensity at \textbf{\textit{Q}} $=$ (0.3,~0, $L$) for the $\Gamma_4$ and $\Gamma_2$ magnetic representations.} 
    \label{fig1}
\end{figure}

\section{results and discussion}
Figure~\ref{fig1} summarizes the XRMS results. The energy scan at \textit{\textbf{Q}} = ($\tau$, 0, 5.5) with $\tau \approx$ 0.3 confirms the presence of antiferromagnetic (AFM) ordering, as this reflection is forbidden for structural Bragg scattering in the tetragonal or orthorhombic structures proposed for this family compounds. A pronounced resonant enhancement is observed at Tb $L_3$ edge, characteristic of magnetic scattering arising primarily from electric dipole ($E$1) transitions between the 2$p$ core level and the unoccupied 5$d$ states.~\cite{KimJW-2005} 

To determine the magnetic moment direction, measurements were performed at \textit{\textbf{Q}} = (0,~$\tau$,~5.5) where $\tau \approx$ 0.3, with the azimuthal angle adjusted so that the [0,~1,~0] direction was aligned within the scattering plane [Fig.~\ref{fig1} (b)]. In this configuration, only the components of the Tb magnetic moments projected onto the scattering plane, namely those along the \textit{\textbf{b}} and \textit{\textbf{c}} directions, contribute to the XRMS intensity. The XRMS intensity can be expressed as $I \propto |F|^2$, where the magnetic structure factor is given by
\[
F \propto \sum_j f^{\mathrm{XRMS}} e^{i\mathbf{Q}\cdot\mathbf{r}_j}.
\]
Here, $f_\mathrm{XRMS} \approx \mathbf{m} \cdot \mathbf{k'}$, where $\mathbf{k'}$ is the scattered wave vector and $\mathbf{m}$ is the magnetic moment direction. The Lorentz factor was included in the calculated intensities to enable a direct comparison with the experimentally measured values.
No magnetic intensity is observed at \textit{\textbf{Q}} = (0,~0.3,~5.5), indicating that the Tb moments do not have components along the \textit{\textbf{b}} or the \textit{\textbf{c}} directions. Instead, the moments are oriented perpendicular to the magnetic propagation vector, which lies along the \textit{\textbf{a}} axis. Consistently, the strong resonant signal at \textit{\textbf{Q}} = ($\tau$, 0, 5.5) with $\tau \approx$ 0.3 [Fig.~\ref{fig1} (a)] arises from magnetic moments aligned along the the \textit{\textbf{b}} direction, which is perpendicular to \textit{\textbf{Q}} in this geometry.

\begin{figure}[!t]
    \centering
    \includegraphics[width=1\linewidth]{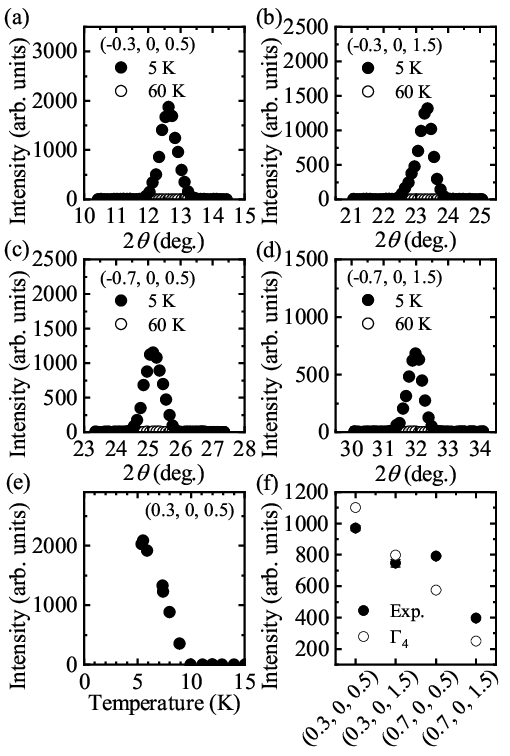}\\
    \caption{Neutron diffraction results. $\theta-2\theta$ scans measured at selected \textbf{\textit{Q}}'s: (a) \textbf{\textit{Q}} $=$ (-0.3,~0,~0.5), (b) (-0.3,~0,~1.5), (c) (-0.7,~0,~0.5), and (d) (-0.7,~0,~1.5). (e) Magnetic order parameter measured at \textbf{\textit{Q}} $=$ (0.3,~0,~0.5). (f) Comparison between the measured intensity and the calculated intensity at various \textit{\textbf{Q}} for the $\Gamma_4$ magnetic representation.}
    \label{fig2}
\end{figure}

The temperature dependence of the magnetic order parameter and the incommensurability are shown in Figs.~\ref{fig1} (c) and (d). The magnetic intensity at \textit{\textbf{Q}} = (0.3, 0, 5.5) emerges below $T_\mathrm{N} =$ 10 K and increases upon cooling, saturating below $T \lesssim 4$ K [Fig.~\ref{fig1} (c)]. The position of the magnetic Bragg peak evolves with temperature, appearing initially at $H \approx 0.306$ just below $T_\mathrm{N}$ and gradually shifting toward $H \approx 0.302$ upon cooling below $T \lesssim 7$ K [Fig.~\ref{fig1} (d)]. Throughout this work, we denote the incommensurate propagation vector as $H = \tau \approx 0.3$ for simplicity.

To determine the stacking of Tb moments along the \textit{\textbf{c}} axis, we measured the $L$ dependence of the magnetic Bragg peaks, as shown in Fig.~\ref{fig1} (e). The magnetic intensity is strongest at half-integer positions with odd integer offsets, indicating a nontrivial magnetic stacking along the \textit{\textbf{c}} direction. The observed $L$ dependence was analyzed using representation analysis\cite{WILLS2000} for the \textit{P4/nmm} space group with propagation vector \textit{\textbf{q}} = (0.3, 0, 0.5). This yields six possible magnetic representations, \mbox{$\Gamma_\textrm{mag}=\Gamma_1~+~2 \Gamma_2~+~\Gamma_3~+~2 \Gamma_4$}.

Given that the Tb moments are aligned along the \textit{\textbf{b}} axis for \textit{\textbf{q}} = (0.3, 0, 0.5), representations with moments along the \textit{\textbf{a}} or \textit{\textbf{c}} directions can be excluded. This reduces the possible solutions to $\Gamma_2$ and $\Gamma_4$, both corresponding to $\mathbf{m} \parallel \mathbf{b}$, but differing in their stacking along the \textit{\textbf{c}} axis. Specifically, $\Gamma_2$ corresponds to a $(+ - - +)$ arrangement, while $\Gamma_4$ corresponds to a $(+ + - -)$ configuration within the magnetic unit cell. These two configurations produce distinct intensity modulations along $L$. By comparing the measured $L$ dependence with calculated XRMS intensities, we find that the data are well described by the $(+ + - -)$ stacking sequence [Fig.~\ref{fig1} (f)], unambiguously identifying the magnetic structure along the \textit{\textbf{c}} axis. Taken together, the Tb moments are oriented perpendicular to the propagation vector (e.g., along the \textit{\textbf{b}} direction for the (0.3, 0, 0.5) propagation vector) and exhibit the $(++--)$ arrangement along the $c$ axis. It is worth noting that the magnetic moment modulation may be either squared-up or sinusoidal. The nature of this modulation can be determined by investigating the presence or absence of higher-order harmonic reflections.

Single-crystal neutron diffraction independently confirms the magnetic structure determined by XRMS, corresponding to $\Gamma_4$ with $\mathbf{m} \parallel \mathbf{b}$. Figure~\ref{fig2} summarizes the neutron diffraction results. $\theta$–$2\theta$ scans were measured at several magnetic Bragg positions, and representative scans are shown in Figs.~\ref{fig2} (a)$-$(d). Well-defined magnetic peaks are observed at the expected incommensurate positions, consistent with the propagation vector \textbf{\textit{q}} = (0.3,~0,~0.5). We compare the measured peak intensities with the calculated intensities for the $\Gamma_4$ magnetic representation in Fig.~\ref{fig2} (f) and find that the $\Gamma_4$ representation successfully reproduces the observed intensity variation, consistent with the XRMS results.

The temperature dependence of the magnetic intensity at the (0.3,~0,~0.5) reflection is shown in Fig.~\ref{fig2}(f). The magnetic peak appears below $\sim 10$ K and grows continuously upon cooling, in good agreement with the ordering temperature and order parameter obtained from XRMS. This consistency establishes that both probes capture the same long-range antiferromagnetic transition. The measured diffraction intensities at the investigated reciprocal-space positions are fully consistent with Tb moments aligned along the \textit{\textbf{b}} axis, modulated by the propagation vector \textbf{\textit{q}} = (0.3,~0,~0.5) and stacked along the \textit{\textbf{c}} axis in a $(++--)$ sequence. The neutron data therefore provide an independent verification of the $\Gamma_4$ magnetic structure inferred from the XRMS azimuthal and $L$-dependent intensity analyses.

We emphasize that this magnetic structure is different from the previously reported complex magnetic ordering of TbCuAs$_2$ proposed from neutron powder diffraction.\cite{Zhao-2017} In contrast to that earlier interpretation, our single-crystal measurements directly resolve a well-defined incommensurate antiferromagnetic structure with a unique moment direction and stacking sequence, thereby removing the ambiguity inherent in powder-averaged data.  This magnetic structure closely resembles that of GdCuAs$_2$\cite{Ashiwini}, except that the incommensurate-to-commensurate lock-in transition observed in GdCuAs$_2$ is absent in TbCuAs$_2$. In contrast, it differs from the magnetic structure of SmCuAs$_2$, which involves a simple doubling of the unit cell along the \textbf{\textit{c}} axis.\cite{Kim-1}.

\begin{figure}[!t]
    \centering
    \includegraphics[width=1\linewidth]{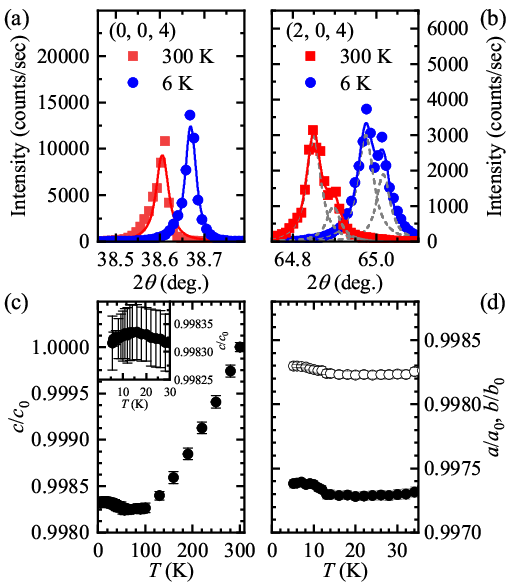}\\
    \caption{Raw longitudinal scans measured at (a) \textbf{\textit{Q}} $=$ (0,~0,~4) and (b) (2,~0,~4) at selected temperatures. Solid lines in (a) and (b) are fit to the data and dashed lines in (b) indicate two-peak fit. (c) Lattice parameter $c$ measured from (0,~0,~4) Bragg peak. The inset shows the variation of the lattice parameters $c$ below 30 K.(d) Lattice parameters $a$ and $b$ obtained from two-peak fit of the data measured at (2,~0,~4) Bragg peak. Solid (open) symbols represent the lattice parameter $a$ ($b$).}
    \label{fig3}
\end{figure}

We now turn to the crystal structure of TbCuAs$_2$. Lattice parameter measurements were performed on the same single crystal used for the XRMS measurements of the antiferromagnetic ordering. The out-of-plane lattice parameter $c$ was determined from the (0,~0,~4) Bragg reflection, while the in-plane lattice parameters were obtained from the (2,~0,~4) reflection. The peak positions were extracted by fitting the diffraction profiles using one or two Lorentzian functions, depending on the observed peak shape. Figures~\ref{fig3} (a) and (b) show the (0,~0,~4) and (2,~0,~4) reflections measured at room temperature ($T$ = 300 K) and $T$ = 6 K. The (0,~0,~4) peak exhibits a slight asymmetry, which may originate from minor misalignment between crystallites or small variations in stoichiometry within the sample volume, and the (2,~0,~4) reflection clearly splits into two peaks. The peak shapes remain consistent at both temperatures, while positions shift systematically with temperature. 

For the tetragonal $P4/nmm$ crystal structure, a single Bragg peak is expected at the (2,~0,~4) position, and the observed splitting therefore indicates a deviation from tetragonal symmetry. A similar splitting has been reported in GdCuAs$_2$, where the (2,~0,~6) reflection separates below $T$ = 60 K and was attributed to a structural distortion to an orthorhombic phase.\cite{Ashiwini} In TbCuAs$_2$, the splitting of the (2,~0,~4) reflection is already present at room temperature and persists to low temperature, indicating that the crystal symmetry is lower than tetragonal over the entire temperature range studied and is consistent with an orthorhombic distortion analogous to that proposed for GdCuAs$_2$.\cite{Ashiwini} 
We also observe that the kink in the lattice parameter $a$ below $T_\mathrm{N}$ is more pronounced than that in $b$, consistent with the behavior reported in GdCuAs$_2$.\cite{Ashiwini} Within a mean-field picture, when lattice distortion couples to antiferromagnetic ordering, lattice contraction occurs along directions with ferromagnetic interactions, whereas lattice expansion occurs along directions with antiferromagnetic interactions.\cite{nematic-1,nematic-2} This is consistent with our observation in TbCuAs$_2$, where $a$ expands while $b$ contracts, reducing the difference between the two and producing a more pronounced kink in $a$.
We note that an earlier study on this family suggested a monoclinic structure as an alternative description, particularly in GdCuP$_{2.2}$,\cite{Mozharivskyj-2000} although the present measurements do not allow us to distinguish between these lower-symmetry structural models.
 
 The temperature dependence of the lattice parameters $c$ and $a$ ($b$) was measured between room temperature (300 K) and the base temperature of the refrigerator ($\sim$ 6 K). The lattice parameter $c$, extracted from the (0,~0,4) reflection, is shown in Fig.~\ref{fig3}(c). Upon cooling, $c$ decreases and reaches a minimum or plateau around $T \approx$ 50 K, followed by a slight increase at lower temperatures and a subsequent decrease below $\sim$ 11 K. This behavior is qualitatively similar to that reported in GdCuAs$_2$,\cite{Ashiwini} although in that case an additional increase in $c$ is observed at the lock-in transition, which is not present in TbCuAs$_2$.

The in-plane lattice parameters $a$ and $b$, determined from the (2,~0,~4) reflection, are shown in Fig.~\ref{fig3}(d). Measurements were performed between $\sim$ 5 K and 35 K. Within the measured temperature range, both $a$ and $b$ decrease slightly or remain nearly constant upon cooling, followed by a slight increase below $\sim$ 11 K. This low-temperature upturn coincides with the onset of long-range antiferromagnetic order and reflects a lattice response to magnetic ordering. The temperature evolution of the in-plane lattice parameters differs from that reported in GdCuAs$_2$,\cite{Ashiwini} where the minimum occurs near $\sim 30$ K and is immediately followed by an increase, which was attributed to a magnetoelastic response associated with the resistivity minimum. Although the origin of this difference remains unclear, the lattice response in TbCuAs$_2$ suggests that the coupling between spin and lattice degrees of freedom may be weaker, becoming pronounced only upon the establishment of long-range magnetic order.

In summary, we have determined the magnetic and structural ground states of TbCuAs$_2$ using XRMS, single-crystal neutron diffraction, and high-resolution X-ray diffraction. The Tb moments order antiferromagnetically below $T_\mathrm{N} \approx$ 10 K with an incommensurate propagation vector  \textit{\textbf{q}} = ($\tau$,~0,~0.5) with $\tau \approx$ 0.3, are aligned along the \textit{\textbf{b}} axis, and adopt a $(++--)$ stacking sequence along the \textit{\textbf{c}} direction corresponding to the $\Gamma_4$ representation. This magnetic structure is unambiguously established by XRMS and independently confirmed by neutron diffraction, resolving the ambiguity of previous powder-based studies. The incommensurate ordering and its weak temperature dependence point to competing exchange interactions and suggest a magnetically frustrated ground state. Structurally, the splitting of the (2,~0,~4) reflection persisting from room temperature to low temperature indicates a deviation from tetragonal $P4/nmm$ symmetry, consistent with a lower-symmetry, likely orthorhombic phase. The temperature evolution of the lattice parameters exhibits anomalies near $T_\mathrm{N}$, indicating coupling between lattice and magnetic degrees of freedom, although the lattice response appears weaker than in GdCuAs$_2$. These results demonstrate that the low-temperature state of TbCuAs$_2$ is governed by coupled incommensurate magnetism and structural distortion, providing a microscopic basis for understanding the interplay of spin, lattice, and electronic behavior in the $R$CuAs$_2$ family.

\begin{acknowledgments}

This work was supported by the University of Wisconsin-Milwaukee. 
This research used resources from the Advanced Photon Source, a U.S. Department of Energy (DOE) Office of Science User Facility operated for the DOE Office of Science by Argonne National Laboratory under Contract No. DE-AC02-06CH11357.
E. D. Mun was supported by the Canada Research Chairs, Natural Sciences and Engineering Research Council of Canada, and Canada Foundation for Innovation program.
Work done at Ames Laboratory (A. S. and P. C. C.) was supported by the U.S. Department of Energy, Office of Basic Energy Science, Division of Materials Sciences and Engineering.  Ames National Laboratory is operated for the U.S. Department of Energy by Iowa State University under Contract No. DE-AC02-07CH11358.
This research used beamline 4-ID of the National Synchrotron Light Source II, a U.S. Department of Energy (DOE) Office of Science User Facility operated for the DOE Office of Science by Brookhaven National Laboratory under Contract No. DE-SC0012704.

\end{acknowledgments}


\bibliographystyle{apsrev4-2-title}
\bibliography{TbCuAs2}

\end{document}